\documentclass[prl,aps,twocolumn,showpacs,superscriptaddress]{revtex4-1}

\usepackage{hyperref}
\hypersetup{colorlinks, citecolor=blue, filecolor= black, linkcolor= black, urlcolor= blue}
\usepackage{graphicx}% Include figure files
\usepackage{dcolumn}% Align table columns on decimal point
\usepackage{bm}% bold math
\usepackage{float}
\begin{document}
\preprint{Preparing}

%=====================  Title   ========================%
\title{Predicted Novel Hydrogen Hydrate Structures under Pressure from First
   Principles}

%=====================  Authors   =====================% 
% GR QIAN
\author{  Guang-Rui Qian}
\email{guangrui.qian@stonybrook.edu/iqianguangrui@gmail.com}
\affiliation{Department of Geosciences, Stony Brook University, Stony Brook, New York 11794-2100, USA}
% AL
\author{ Andriy O. Lyakhov}
\affiliation{Department of Geosciences, Stony Brook University, Stony Brook, New York 11794-2100, USA}
% QZ
\author{Qiang Zhu}
\affiliation{Department of Geosciences, Stony Brook University, Stony Brook, New York 11794-2100, USA}
% Artem
\author{ Artem R. Oganov}  
\affiliation{ Department of Geosciences, Center for Materials by Design, and Institute for Advanced Computational Science, State University of New York, Stony Brook, NY 11794-2100}
\affiliation{
Moscow Institute of Physics and Technology, 9 Institutskiy lane, Dolgoprudny city, Moscow Region 141700, Russia}
\affiliation{School of Materials Science, Northwestern Polytechnical University, Xi'an 710072, China}
% XD
\author{ Xiao Dong }
\affiliation{Department of Geosciences, Stony Brook University, Stony Brook, New York 11794-2100, USA}
\affiliation{School of Physics and MOE Key Laboratory of Weak-Light Nonlinear Photonics, Nankai University, Tianjin 300071, China}

\date{\today}

%=====================  Abstract   =====================% 
\begin{abstract}
\noindent  
Gas hydrates are systems of prime importance. In particular, 
hydrogen hydrates are potential materials of 
icy satellites and comets, and may be used for hydrogen storage.
We explore the H$_2$O$-$H$_2$ system at pressures in the range
0$-$100 GPa with $ab$ $initio$  variable-composition evolutionary
simulations. According to our calculation and previous experiments, 
the H$_2$O$-$H$_2$ system undergoes a series of transformations
with pressure, and adopts the known open-network clathrate 
structures (sII, C$_0$), dense ``filled ice'' structures (C$_1$, C$_2$) 
and two novel hydrate phases. One of these is 
based on the hexagonal ice framework and has the same  
H$_2$O$:$H$_2$  ratio (2:1) as the C$_0$ phase at low pressures and 
similar enthalpy (we name this phase $Ih$-C$_0$).  The other newly 
predicted hydrate phase has a 1:2  H$_2$O$:$H$_2$  ratio and 
structure based on cubic ice. This phase (which we name C3) is 
predicted to be thermodynamically stable above 38 GPa when  
including van der Waals interactions and zero-point vibrational 
energy, and explains previously mysterious experimental X-ray diffraction and Raman measurements. 
This is the hydrogen-richest hydrate and this phase has a
remarkable gravimetric density (18 wt.$\%$) of easily extractable hydrogen.
 
\end{abstract}

%% pacs
%  62.50.-p High-pressure effects in solids and liquids
%  61.66.Fn Inorganic compounds
%  61.50.Lt Crystal binding; cohesive energy
%  63.20.-e Phonons in crystal lattices
%  64.70.K± Solid-solid transitions
%  61.50.Ah Theory of crystal structure, crystal symmetry; calculations and modeling

%\keywords{}
\maketitle
\newpage
%\begin{multicols}

%=====================  Main Context   

%\section{Introduction}

Molecular compounds (cocrystals) of water ice (H$_2$O) and 
hydrogen (H$_2$) are known to form clathrate structures 
with the hydrogen molecules encapsulated as guests in the 
host sublattice formed by water molecules.
Hydrogen hydrates, as environmentally clean and efficient 
hydrogen storage materials, have excited significant interest.
Extensive literature exists from both experimental 
\cite{Vos-PRL-1993, Vos-CPL-1996, Mao-SCI-2002, Mao-PNAS-2004, 
Lokshin-PRL-2004, Hirai-JPCC-2007, Machida-JCP-2008, 
Machida-JPCS-2010, Machida-PRB-2011, Hirai-JCP-2012, Timothy-JPCC-2011, 
Efimchenko-JAllCom-2011, Timothy-preparing}  and 
theoretical \cite{Lenz-JPCA-2011, Zhang-JCP-2012} sides. 
Aside from the H$_2$ molecules, 
many other small molecules are known to form clathrate 
structures as guest species under elevated pressure as well, 
including noble gases, nitrogen, oxygen, methane 
etc. (See Ref. \cite{Loveday-PCCP-2008} and references therein) 
Hydrogen hydrates are important as potentially major 
materials of icy satellites and  comets, and potential hydrogen 
storage materials.

Twenty years after the first report of  the formation of two 
filled-ice hydrogen hydrates by Vos $et$ $al.$ 
\cite{Vos-PRL-1993},  four hydrogen hydrate 
forms are known to exist at elevated 
pressures. Two of the hydrogen hydrates are clathrates, denoted as
clathrate structure II (sII) \cite{Mao-SCI-2002, Lokshin-PRL-2004} 
and compound 0 (C$_0$) 
\cite{Efimchenko-JAllCom-2011, Timothy-preparing}, 
the other two are filled ice hydrates, compound 1 (C$_1$) 
and compound 2 (C$_2$) \cite{Vos-PRL-1993, Vos-CPL-1996}. 
The sII clathrate hydrate was synthesized under pressures of 
180 to 220 MPa at 300 K, and its structure was shown to contain 
48 hydrogen molecules and 136 
water molecules in the unit cell \cite{Mao-SCI-2002}.
The C$_0$ clathrate was recently found to be stable near 
0.5 GPa and to have the composition 2H$_2$O$:$1H$_2$ and 
trigonal structure \cite{Efimchenko-JAllCom-2011}. 
The water molecules in the C$_0$ structure are arranged in a 
totally new way, different from the known ices or ice 
sublattices in hydrates structures. This structure has space 
group $P3_221$, but this could possibly go as low as $P3_2$, 
depending on how the hydrogens are arranged 
\cite{Efimchenko-JAllCom-2011,Timothy-preparing}.

At higher pressures, clathrates give way to denser structures of the
filled ice type. The C$_1$ and C$_2$ phases are formed
at 0.36$-$0.9 GPa and $\sim$2.4 GPa, respectively 
\cite{Vos-PRL-1993, Vos-CPL-1996,  Timothy-JPCC-2011}. 
The C$_1$ hydrate has a water host framework based on 
ice-II and a 6:1 water to hydrogen ratio. C$_2$ has a 1$:$1 ratio of 
water to hydrogen and is composed 
of water molecules in the ``cubic ice'' (ice-Ic) framework
and rotationally disordered hydrogen molecules 
\cite{Loveday-PCCP-2008}. Recent experiments 
\cite{Machida-JCP-2008, Machida-JPCS-2010, Machida-PRB-2011, Hirai-JCP-2012} 
indicate that the C$_2$ hydrate undergoes a structural 
transformation from cubic to tetragonal phase at around 10-20 
GPa, with an increasing difference in the unit cell axes, 
and then transforms to another high-pressure phase near 
$\sim$45 GPa. This high-pressure phase is maintained up to 
at least 80 GPa but its structure is not fully resolved.
Given the difficulties in characterization of the chemical 
composition and crystal structure of these hydrates, and 
believing that new phases are likely to exist, we decided 
to perform a computational search to revisit the 
H$_2$O$-$H$_2$ system under pressure.

Using the evolutionary algorithm USPEX 
\cite{USPEX-2010-RevMinGeo, Oganov-ACR-2011, USPEX-mole-2012,
XeO-NatureC-2012}, we explored all possible stable phases in the 
H$_2$O$-$H$_2$ system. Predictions were done in the 
variable-composition mode at several pressures 
(0, 1, 2, 5, 10, 20, 50 and 100 GPa) and zero temperature. 
A number of studies illustrate the power of the USPEX 
method (for example, \cite{CHH-PRL-2013, QZ-PCCP-2013, WWZhang-Science-2013}). 
Given molecular nature of all stable and nearly stable compounds 
in this system, we searched for the packing of 
well-defined H$_2$O and H$_2$ 
molecules (rather than H and O atoms), by applying the
specially designed constrained global optimization algorithm
\cite{Method-Intro}, considering structures containing 
up to 24 molecules (i.e. up to 72 atoms) per primitive unit cell.

Structure relaxations were done using density functional theory 
(DFT) within  van der Waals (vdW)  functional
optB88-vdW \cite{VASP-VDW-2011} in the framework of the all-electron 
projector augmented wave (PAW) \cite{PAW-1994} 
method as implemented in the VASP \cite{VASP-PRB-1996} code. 
The plane wave kinetic energy cutoff of 600 eV and Monkhorst-Pack 
$k$-point \cite{Kpoint-PRB-1991} meshes with the reciprocal space
resolution of 2$\pi$$\times$0.05 $\AA$ were used. Having identified 
the most stable compositions and candidate structures, we 
relaxed them at pressures from 1 $atm$ to 120 GPa with an 
even higher cutoff of 800 eV to refine their thermodynamic 
properties and stability fields. Structure relaxations proceeded until net forces on atoms were below 
1 meV/$\AA$, which gave us enthalpies converged to better than 1 meV/atom. 

It is expected that the relative contribution of hydrogen bonding 
(H-bonding)  and van der Waals (vdW) dispersion forces has 
a significant impact on the phase transition pressures 
and cohesive properties of the various crystalline ice phases 
\cite{iceVDW-PRL-2011}. This is also confirmed by our calculations
(see the Supplemental Material \cite{Supplemental} for the 
phase transition pressures of ice phases from optB88-vdW, 
GGA \cite{GGA-1996} calculations and experiments). 
Thus, all calculations included the
vdW functional to treat the vdW forces, unless stated otherwise.

 \begin{figure}[htb!]
\centerline{\includegraphics[width=8.5cm]{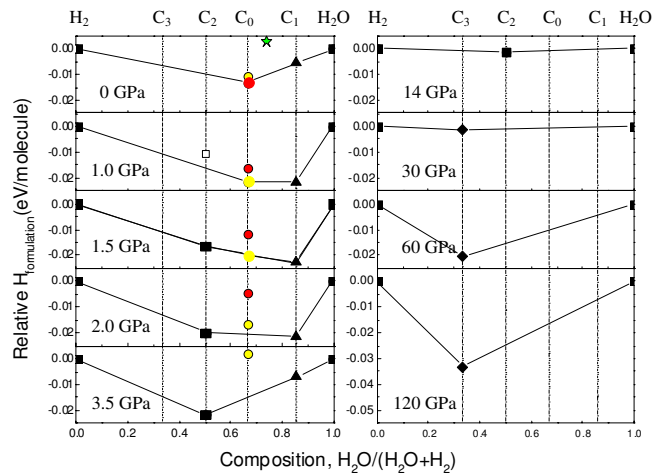}} 
\caption{(Color online) Convex hull diagram for H$_2$O-H$_2$ 
system at selected pressures and zero temperature.   
This figure shows the enthalpy of formation (in eV/molecule) of 
molecular compounds from H$_2$O and H$_2$. The 
red and yellow circles represent the C$_0$ and $Ih$-C$_0$ phases, respectively.
The green star represents the sII structure.
}
\end{figure}

Remarkably, we have found two novel filled ice hydrogen 
hydrates, and all known hydrogen hydrates (except the sII structure, 
because of the very large number of molecules in its unit cell). 
Thus, at pressures in the range 0$-$2 GPa, the sII structure is input 
separately in order to calculate stability ranges of phases in the 
H$_2$O-H$_2$ system. Fig. 1 shows the convex hull diagram for the 
H$_2$O-H$_2$ system.

Our results are in generally very good agreement with experiments, 
but with several novel aspects. At 0 GPa,  the C$_0$, C$_1$ and 
a novel hydrogen hydrate phase are found stable or nearly stable 
in the H$_2$O$-$H$_2$ system, while the sII phase is metastable 
($\sim$0.013 eV/molecule less stable than the mixture of stable 
compounds C$_0$ and C$_1$).  
The structure of the novel hydrogen hydrate is based on the 
framework of hexagonal ice (ice-Ih), with two hydrogen molecules 
hosted inside channels running along the hexagonal axis (Fig. 2a). 
It has a 2:1 ratio of water to hydrogen, same as 
C$_0$, and has space group $Cc$. We name it $Ih$-C$_0$ to 
distinguish from C$_0$. The enthalpy  of the $Ih$-C$_0$  phase is 
close to C$_0$, and is slightly lower at pressures above
$\sim$0.4 GPa (see the Supplementary Material \cite{Supplemental}).
At 1.5 GPa, in addition to  the C$_0$, $Ih$-C$_0$ and C$_1$ phases, 
the hydrate phase C$_2$ with an ice-Ic framework structure 
becomes stable. 

 \begin{figure}[htb]
\centerline{\includegraphics[width=8.5cm]{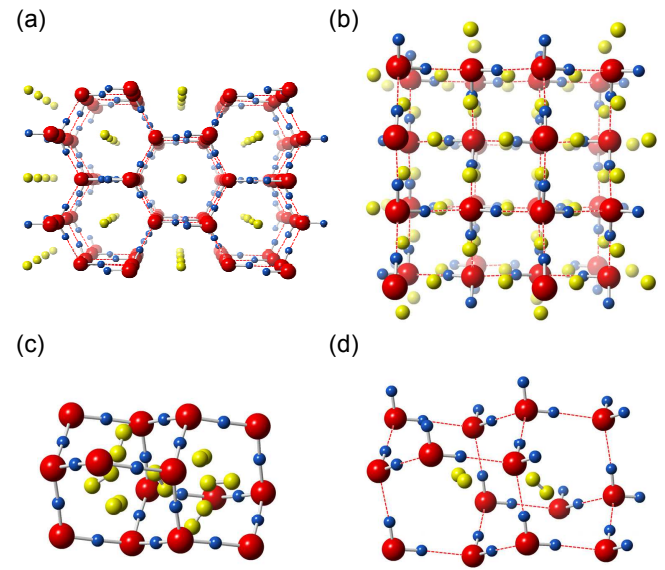}} 
\caption{(Color online) 
(a) Hydrate $Ih$-C$_0$ structure at 0.5 GPa, 
(b) hydrate C$_3$ structure at 30 GPa, 
(c) cages formed by water molecules in hydrate C$_3$ at 
100 GPa, the hydrogen molecules are located at the center 
of each chair-like H-O ring,
(d) cages in ``filled ice-Ic'' hydrate C$_2$, hydrogen molecules 
are  in the center of the cage. Large red and small blue spheres 
are O and H atoms in water molecules, respectively; 
the yellow spheres represent the H$_2$ molecules in (a) and (b), 
and represent H atoms in (c) and (d).
Red dashed lines represent hydrogen bonds.
}
\end{figure}

At pressures above 2 GPa, the C$_0$ and $Ih$-C$_0$ phases are 
calculated to be above the convex hull, indicating that these 
phases become unstable against decomposition into 
C$_1$ and C$_2$. Above 3.5 GPa, the C$_1$ phase  will also 
become unstable, and the C$_2$ phase will remain 
the only stable hydrate. For hydrate C$_2$,  USPEX calculations uncovered at least four 
typical energetically favorable candidate structures \cite{H4O} at different 
pressures, $P4_12_12$, $I4_1/amd$, $Pna2_1$ and $I4_1md$ 
(see the Supplemental Material \cite{Supplemental}), 
which is similar to Ref. \cite{Zhang-JCP-2012}. 
The C$_2$ phase will lose stability at $\sim$14 GPa, which is much lower than 
40 GPa suggested in the previous study \cite{Machida-JCP-2008, 
Hirai-JCP-2012}. We explain this by metastable persistence of 
C$_2$ up to the pressure of 40 GPa. Between 14$-$28 GPa,  there 
are, unexpectedly, no thermodynamically stable hydrates.

Near 30 GPa, another novel H$_2$O$-$H$_2$ phase is found to be
stable at zero temperature. It has a 1:2 water to hydrogen ratio,
and net composition H$_6$O. 
This novel hydrogen hydrate, which we name C$_3$, has the 
highest hydrogen concentration among all hydrogen hydrates. 
If it can be synthesized at low pressures, it would be an attractive 
hydrogen storage material, having 18 wt.$\%$ concentration of 
easily separable (non-water) hydrogen.
The C$_3$ structure has space group $P4_1$ and is also based 
on the framework of ice-Ic (Fig. 2b),  similar to low-pressure 
hydrate C$_2$. The unit cell of C$_3$ contains four 
water molecules, the H$_2$ molecules are located  
at the center of chair-like H-O rings (formed by six oxygen and 
six hydrogen atoms) that form faces of the cage, as 
shown in Fig. 2c. Differently, in the C$_2$ 
hydrate,  the H$_2$ molecules are in the center of the water 
cages (Fig. 2d). According to our calcualtions, the C$_3$ phase will remain stable 
up to at least 120 GPa.

Our theoretical calculations  indicate that the H$_2$O$-$H$_2$ 
system contains several stable phases,  
including open-network clathrate structures (C$_0$) and dense
filled ice phases ($Ih$-C$_0$, C$_1$, C$_2$ and C$_3$). 
The C$_0$ phase is predicted to be stable at pressures below 
1.5 GPa, which is close to the experiments result 
(below 0.8 GPa \cite{Efimchenko-JAllCom-2011}). The C$_1$ phase
is predicted to be stable at pressures below 3.5 GPa, 
also close  to  the experimentally determined transition pressure of 
2.5 GPa \cite{Vos-PRL-1993}. 
The zero-point vibration energy (ZPE)  significantly affects the 
relative stability of hydrogen-rich structures 
\cite{Pickard-Nature-2007}. We have estimated the ZPE 
within the quasi-harmonic approximation \cite{PHONOPY-PRB-2008} 
to refine the stability ranges of C$_2$ and C$_3$ phases above 
10 GPa. When considering the ZPE, the stability field of 
the C$_2$ phase expands up to $\sim$19 GPa, but this phase 
remains dynamically stable, and thus can exist as a 
metastable material at pressures of at least 60 GPa 
(see the Supplementary Material \cite{Supplemental}).

The C$_3$ phase starts to be energetically favorable above $\sim$38 GPa 
when including ZPE, as shown in Fig. 3. Thus, the novel C$_3$ phase 
can be synthesized in hydrogen-rich conditions at pressures starting 
from 38 GPa. This theoretical value agrees well with the transition 
pressure 45$-$50 GPa to the hitherto mysterious phase of 
unknown composition \cite{Machida-JCP-2008, Hirai-JCP-2012}.
As shown in Fig. 4, the Raman shift calculations \cite{Lazzeri-PRL-2003} 
reveal the the H$_2$ vibron Raman shift differences between the C$_2$ and C$_3$ 
phases in H$_2$-D$_2$O system. The Raman shift of C$_3$ phase, 
rather than an amorphous phase, agrees very well with
the lower Raman frequencies of the vibron for the hydrogen 
molecules observed in Ref. \cite{Machida-PRB-2011}. The black 
rhombi in Fig. 4 indicate that some of the H$_2$-D$_2$O C$_3$ sample 
encountered decomposition when quenched to low pressure. The variation 
of lattice parameters of the ice host structure in hydrates with pressure, revealed 
by our theoretical calculations, also agrees well with the observation 
from the XRD results at high pressure \cite{Hirai-JCP-2012}.
At 55 GPa, our calculation gives lattice parameter of C$_3$
phase $a$=$b$=4.00 $\AA$ and $c$=5.67 $\AA$, corresponding to cubic ice sublattice 
with periodicity 5.67 $\AA$, where experiments gives $\sim$5.5 $\AA$ \cite{Hirai-JCP-2012}.
At low pressure, the C$_2$ adopts a ``cubic ice'' host structure and then 
transforms to a ``tetragonal'' one around 20 GPa \cite{Hirai-JCP-2012} 
(see the Supplementary Material \cite{Supplemental} for a comparison). 
When forming the C$_3$ phase at increased pressure and 
in excess of H$_2$, the ice host structure transforms 
to the ``cubic ice'' again.  The change from tetragonal to ``cubic'' structure occurs before 
H-bond symmetrization transition happens in ``tetragonal'' 
type C$_2$ around 55 GPa. Thus, such structural transformation is unrelated to 
symmetrization of the H-bonds, but comes from the emergence of 
the C$_3$ phase.  For the hydrate C$_3$, the H-bond 
symmetrization is predicted to occur at $\sim$120 GPa (see the Supplementary Material \cite{Supplemental}),
which is close the theoretical H-bond symmetrization 
pressure in ice-VII \cite{Zhang-JCP-2012}. 

 \begin{figure}[htb]
\centerline{\includegraphics[width=8.5cm]{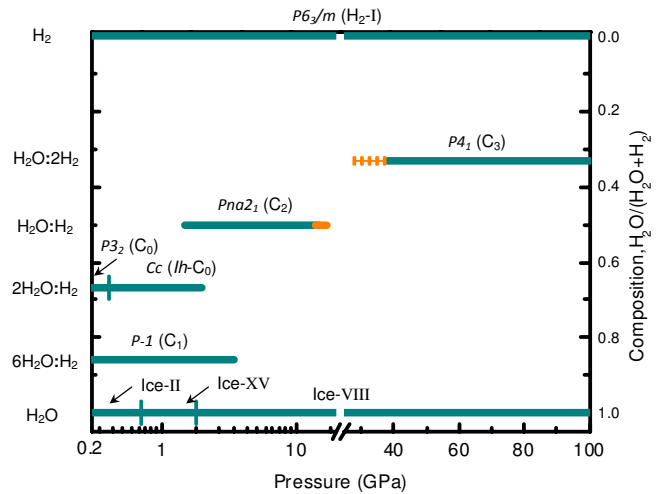}} 
\caption{(Color online) 
Phase diagram of the H$_2$O$-$H$_2$ system. The stability ranges 
of C$_2$ and C$_3$ phases are calculated with and without 
ZPE effect. The solid orange line represents extra stability range 
added due to ZPE, the dashed orange line represents regions 
that become unstable after inclusion of the ZPE.
}
\end{figure}

 \begin{figure}[htb]
\centerline{\includegraphics[width=8.5cm]{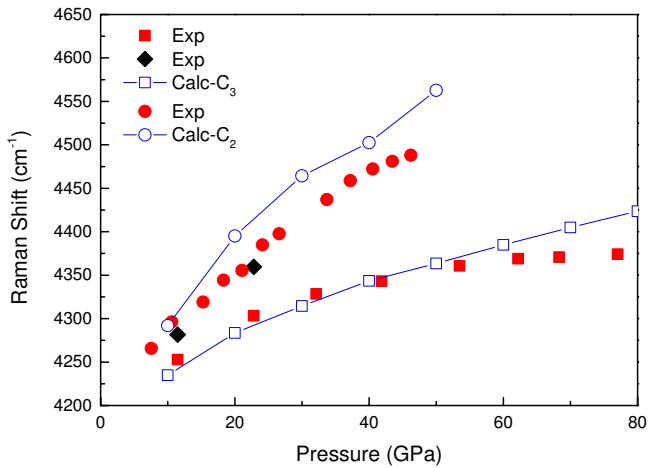}} 
\caption{ (Color online) 
Variations of the Raman shift of the vibron for the H$_2$ molecules with pressure
from experimental data in \cite{Machida-PRB-2011} and our
theoretical calculations.
The red and black symbols are the experimental data for  H$_2$ vibrons in 
the H$_2$-D$_2$O sample. The blue open circles  and squares indicate the 
Raman shift calculation for C$_2$ and C$_3$ phases of
H$_2$-D$_2$O system, respectively.
}
\end{figure}

The C$_2$ and C$_3$ hydrates have a similar ice host framework, 
but the different numbers of hydrogen molecules and their 
different locations and orientations bring huge differences in phase 
stability range. In the C$_2$ phase, hydrogen molecules stay in the 
centers of cages formed by water molecules $Ð$ in contrast to 
C$_3$ phase, where they are located at the faces of the cages. 
To clarify the causes of stability of hydrogen hydrates, we used 
Bader analysis \cite{Bader-1990,Henkelman-CMS-2006}, and 
focused on the C$_2$ and C$_3$ phases (see the Fig. S6 in Supplementary Material \cite{Supplemental}). 
We found a very small charge transferred from H$_2$ to
water molecules, so that the H$_2$ molecules are slightly positively 
charged, and H$_2$O molecules carry a slight negative charge. 
The magnitudes of these charges are $\sim$10$^{-3}$$-$10$^{-2}$ 
per molecule.  This suggests that interactions between 
these molecules are almost purely steric, 
mainly related to packing density and shapes of the molecules.
Comparing Bader volumes of the H$_2$O and H$_2$ molecules in 
the hydrates and in pure H$_2$O and H$_2$, we see that water 
molecules occupy slightly larger volume in the hydrates, whereas 
hydrogen molecules occupy much less space in C$_3$ hydrate 
than in pure H$_2$ $-$ this leads to net densification, stabilizing 
this phase in a wide pressure range. 
For the C$_2$ hydrate, the H$_2$ molecules have lower volume 
than in pure H$_2$ only at pressures below $\sim$10 GPa, 
which explains its instability at higher pressures.

 \begin{figure}[htb]
\centerline{\includegraphics[width=8.5cm]{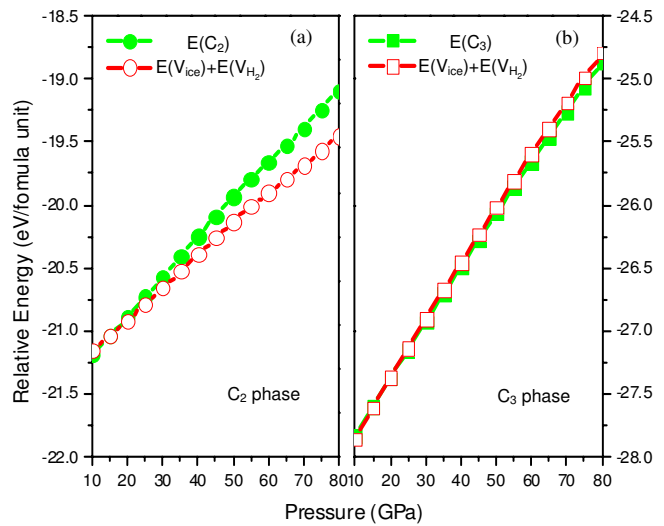}} 
\caption{ (Color online) 
Internal energy of the C$_2$ and C$_3$ phases relative
to the isochoric mixture of H$_2$O and H$_2$. Green lines 
represent the energy of the hydrate phases; red lines $-$ 
the energy of the isochoric mixture of ice-VIII and 
H$_2$-I phases.
}
\end{figure}

Having considered the PV-term in the enthalpy ($H=E+PV$),
to get additional insight, we turned to the internal energy E
 and its changes when the H$_2$ and H$_2$O 
molecules are placed from the hydrate into pure H$_2$ and 
H$_2$O phases, while keeping molecular volumes fixed to their 
values in the hydrate (Fig. 5). This energy characterizes the 
net balance between the vdW attraction and 
steric repulsion between the molecules: this net effect is very 
small in the C$_3$ phase (slightly destabilizing below 
$\sim$30 GPa and slightly stabilizing above $\sim$30 GPa). The remarkably 
wide stability field of the C$_3$ phase is therefore mostly due to its 
high density and only to a small extent to more favorable 
intermolecular interactions. A much more interesting picture is 
observed for the C$_2$ phase (Fig. 5a): we find its slight energetic 
stabilization below $\sim$15 GPa, and an increasingly large destabilization 
at higher pressures. This explains why C$_2$ is unstable at high 
pressures, and furthermore, it is clear that the increasing energetic 
instability of the C$_2$ phase is responsible for the displacive phase 
transition, metastably occurring on overcompression and 
transforming the cubic H$_2$O host sublattice into tetragonal, 
to enable better packing of the molecules. 

Our calculations found that a C$_3$-type phase is stable in the H$_2$O$-$He system at 
8-75 GPa (without including zero-point energy), and this phase is denser than 
the mixture of H$_2$O and He. On the other hand, no such phase was found 
in the H$_2$O-Ne system, and indeed the C$_3$ phase is not 
packing-efficient in this system (see the Fig. S7 and Fig. S8 in 
Supplementary Material \cite{Supplemental}). 
He and Ne are equally chemically inert, their almost only differences are size and 
(here insignificant) mass. Stability of He-C$_3$ and instability of Ne-C$_3$ hydrates reinforce
 our conclusion made for the H$_2$O$-$H$_2$ system, that stability of this novel 
phase comes not from specific bonding 
interactions between the molecules, and not even due to their 
shapes, but mostly due to their very efficient packing.

In summary, using the evolutionary algorithm USPEX, we 
explored the H$_2$O$-$H$_2$ system at pressures of up to 100 GPa. 
Stoichiometries and stability fields of H$_2$O$-$H$_2$ hydrate phases 
have been studied. A series of pressure-induced transformations 
found by theory closely 
coincides with experimental data, but also new insight was obtained. 
A novel $Ih$$-$C$_0$ structure is predicted to have a very close 
enthalpy to the recently discovered C$_0$ structure. At pressures 
above 38 GPa, novel hydrogen hydrate C$_3$, based on cubic 
ice Ic, is predicted to be stable. With stoichiometry H$_2$O$:$2H$_2$, 
this is the hydrogen-richest hydrate known to date. With gravimetric 
density of easily removable hydrogen (18 wt.$\%$), this is a promising 
hydrogen storage material that can find practical applications
 if its synthesis pressure can be decreased.

We thank the DARPA (Grants No. W31P4Q1310005 
and No. W31P4Q1210008), National Science Foundation 
(EAR-1114313, DMR-1231586) for financial support.

We thank  Purdue University Teragrid and TACC Stampede system
for providing computational resources and technical support for this work 
(Charge No.: TG-DMR110058).

\bibliography{projectHO} 

\begin{thebibliography}{37}
\bibitem [{Met()}]{Method-Intro}%
  \BibitemOpen
  \href@noop {} {}\bibinfo {note} {We have done two types of
  variable-composition structure predictions in searching for all stable phases
  in the {H}$-${O} system: (1) {I}n the {H}$-${O} system, assembling the
  structures from atoms, and (2) {I}n the {H}$_2${O}$-${H}$_2$ system with
  giving {H}$_2${O} and {H}$_2$ molecules as structure building blocks. {W}e
  have found that, even in the {H}$-${O} system, the stable stats at pressures
  of our interesting ($\textless$120 {G}{P}a) are actually using made of
  well-defined {H}$_2${O} and {H}$_2$ molecules. This allowed us to focus on
  molecular-type calculations, capable of efficiently dealing with large
  systems, without loss of rigor.}\BibitemShut {Stop}%
\end{thebibliography}
\bibliographystyle{apsrev4-1}

%---------------------------------------------------------------------------------------------

\end{document}